\documentclass{PoS}
\usepackage{amsmath}
\usepackage{mathrsfs}
\usepackage[utf8]{inputenc}

\def\bec{\begin{center}}
\def\eec{\end{center}}
\def\beq{\begin{equation}}
\def\eeq{\end{equation}}
\def\bea{\begin{eqnarray}}
\def\eea{\end{eqnarray}}
\def\meas{\int d^{4}x \sqrt{-g}}

\def\B{\Box}

\def\xy{\mathrm{xy}}

\def\x{\mathrm{x}}
\def\y{\mathrm{y}}
\def\cos{\mathrm{cos}}

\title{Lattice quantum gravity with scalar fields}
\ShortTitle{Lattice quantum gravity with scalar fields}
\author{\speaker{Raghav G. Jha}, Jack Laiho, Judah Unmuth-Yockey \\
  Department of Physics, Syracuse University, Syracuse, NY 13244, USA \\
  E-mail: \email{rgjha@syr.edu} \\
}

\abstract{ 
We consider the four-dimensional Euclidean dynamical triangulations lattice model of quantum gravity based on
triangulations of $S^{4}$. We couple it minimally 
to a scalar field in the quenched approximation. Our results suggest a multiplicative renormalization for
the mass of the scalar field which is consistent with the shift symmetry of the discretized lattice
action. We discuss the possibility of measuring the mass anomalous dimension and the gravitational
binding energy between two scalar test particles, where a negative bound
state energy would imply that this model has an attractive gravitational force. 
}

\FullConference{The 36th Annual International Symposium on Lattice Field Theory \\  July 22-28, 2018 \\ 
Michigan State University, East Lansing, Michigan, USA  } 

\begin{document}
\setlength{\abovedisplayskip}{6 pt}
\setlength{\belowdisplayskip}{6 pt}

\section{Introduction} 
The problem of constructing a quantum theory of gravity in four dimensions is notoriously difficult because the 
theory is perturbatively non-renormalizable in four dimensions. 
However, it was conjectured by Weinberg in \cite{Weinberg:1980gg} that the quantum theory of gravity might be 
\textit{asymptotically safe} in the ultraviolet (UV) by possessing an 
interacting fixed point with finite dimensional critical surface. There is no proof that this is true, but there
is non-trivial evidence in its support ; see \cite{Litim:2003vp, Niedermaier:2006ns} and references 
therein. However, there are arguments that argue that gravity cannot be described by a renormalizable quantum 
field theory in four dimensions based on the inconsistency between entropy scaling of 
black holes in the high-energy spectrum and that of a scale invariant theory 
~\cite{Aharony:1998tt, Shomer:2007vq}. The argument goes that the entropy of a renormalizable theory 
must scale as $S \sim E^{\frac{d-1}{d}}$. For gravity, the high-energy spectrum must be dominated
by black holes. The black hole entropy area law tells us that $ S \sim E^{\frac{d-2}{d-3}}$.
However, these scalings are consistent for the value $d = 3/2$, which curiously is also
 the value of the fractal (spectral) dimension obtained from lattice gravity simulations at 
 short distances in both Euclidean dynamical triangulations (EDT) and Causal dynamical triangulations (CDT) 
\cite{Laiho:2011ya, Coumbe:2014noa}. 

In this proceedings, we work under the assumption that there is merit in understanding gravity as a quantum field theory 
in four dimensions. In the 1990s, using the discretization of the Einstein-Hilbert action due to Regge, 
the phase diagram of dynamical triangulations was explored and it was hoped that one would find a second-order critical point, which could be 
identified as the asymptotically safe fixed point of Weinberg's conjecture. This is crucial for the continuum extrapolation
of the lattice model. However, the order of the phase transition turned out to be 
first-order ~\cite{Bialas:1996wu, deBakker:1996zx}. The approach we follow here introduces a non-trivial measure 
term which alters the structure of the phase diagram where it appears that the continuum limit can be taken. 

\section{Action and discretization}
The Einstein-Hilbert action for a metric $g_{\mu\nu}$ has the form,  
\beq
S_{E}[g] = \frac{1}{16\pi G} \meas \left ( 2 \Lambda - R \right) 
\eeq
where R is the Ricci scalar and $\Lambda$ and G are the cosmological constant and Newton's constant respectively. 
On a triangulation, we discretize the volume according to 
\beq
V_{4} = \meas  \rightarrow N_{4}[T] 
\eeq
The Euclidean Einstein-Regge discrete action is given by :

\beq\label{eq:lattice_act}
S_{ER} = -\kappa_{2} N_{2} + \kappa_{4} N_{4} 
\eeq
where $N_{i}$ is the number of simplices of dimension $i$, $\kappa_{2}$ and $\kappa_{4}$ are related 
to Newton's constant $G$ and the cosmological constant $\Lambda$. In our lattice simulations, 
$\kappa_{4}$ must be tuned to a critical value such that an infinite lattice volume can be taken. This leaves two parameters in the theory, $\kappa_{2}$ 
and $\beta$. Note that for a fixed topology $S^{4}$ with fixed four-volume $N_{4}$, the number of 0-simplices and 2-simplices 
$\textit{i.e}$ $N_{0}$ and $N_{2}$ are not independent but related by

\[ N_{2} = 2N_{0} + 2 N_{4} - 2\chi ~~, \] 
where $\chi = N_{0} - N_{1} + N_{2} - N_{3} + N_{4} $ is the Euler characteristic. 
It is customary to start with the discrete Euclidean-Regge action \cite{Regge:1961px}  

\beq 
\label{eq:SER}
S_{ER} = -\kappa \sum V_{2} \left( 2\pi - \sum \theta \right) + \lambda \sum V_{4} 
\eeq
where $\kappa = \frac{1}{8\pi G}$, $\theta = \cos^{-1}(1/4)$ and $ \lambda = \kappa \Lambda $. Also, 
the volume of a d-simplex is given by :

\begin{equation}
\label{eq:volume}
V_{d} = l^{d} \frac{\sqrt{d+1}}{\sqrt{2^{d}} d !}. 
\end{equation}
Rewriting $V_{2}$ and $V_{4}$ above in terms of $N_{2}$ and $N_{4}$ and using the relation 
between them ($ N_{2} = 10 N_{4}$) and using Eq.~\eqref{eq:volume}, we can write Eq.~\eqref{eq:SER} as, 

\begin{equation}
S_{ER} = - \frac{\sqrt{3}}{2} \pi \kappa N_{2} + N_{4} \left ( \kappa \frac{5 \sqrt{3}}{2} \cos^{-1} \left(\frac{1}{4}\right) +\frac{\sqrt{5}}{96} \lambda \right ) 
\end{equation}
Defining new variables $\kappa_{2} = \frac{\sqrt{3}}{2} \pi \kappa $ 
and $\kappa_{4}$ = $\kappa$ $\frac{5 \sqrt{3}}{2} \cos^{-1} (\frac{1}{4}) +\frac{\sqrt{5}}{96}$ $\lambda $,  
we recover the form written above in Eq.~\eqref{eq:lattice_act}. 
The continuum path integral can be written as, 
\begin{equation}
Z = \int \mathscr{D}[g] e^{-S_{ER}[g]}
\end{equation}
which we discretize by writing the grand canonical ($\textit{i.e.}$ volume can fluctuate) partition function as, 
\begin{equation}
Z = \sum_{\mathcal{T}} \frac{1}{C(\mathcal{T})}\Bigg[ \prod_{i=1}^{N_{2}} \mathcal{O}(t_{i})^{\beta} \Bigg] e^{-S_{ER}[\mathcal{T}]}
\end{equation}
where $C(\mathcal{T})$ is a symmetry factor which mods out the number of equivalent ways of labelling
the vertices in a given triangulation $\mathcal{T}$. Note that the term in paranthesis is the contribution from 
a non-trivial measure term. It was shown \cite{Bruegmann:1992jk} that this term is crucial in four 
dimensions unlike in lower dimensions.
This term was absent in much of past work \cite{deBakker:1996qf} which implicitly assumed $\beta=0$ triangulations. 
Another difference from what was done in the past is that we use \textit{degenerate triangulations} \cite{Bilke:1998bn} as opposed to the 
combinatorial ones used in \cite{deBakker:1996qf}. 
The reason we employ degenerate triangulations is that it leads to a substantial reduction in finite-size effects compared to
combinatorial triangulations. 

\section{Coupling to the scalar field} 
We introduce the scalar field $\Phi$ as a matter field minimally coupled to gravity. In the continuum,
the action coupled to the scalar field ignoring the self-interaction terms can be written as
\beq
S = S_{ER}[g]  +  S[g, \Phi] 
\eeq
where,
\beq
S[g,\Phi] = \meas ~ \bigg ( g^{\mu\nu} \partial_{\mu}\Phi \partial_{\nu}\Phi + m_{0}^2 \Phi^2 \bigg) 
\eeq
Here, $\Phi$ is a test particle and the back reaction of the metric is ignored. 
The notion of distance is non-trivial to define on the lattice that itself is dynamical, but the 
geodesic distance can be understood via the propagation
of massive particle as suggested in ~\cite{1992NuPhB.368..671D, Hamber:1993gn} 
The propagator for the scalar field in a \emph{fixed background} of constant curvature decays as,  
\beq
\label{eq:prop}
G_{\xy}(r) = A(r) e^{-mr} 
\eeq
where $r$ is the geodesic distance between two points. It is noted that $m$, is the physical mass 
of the particle with
\emph{multiplicative gravitational renormalization}.  The multiplicative renormalization follows from the shift 
symmetry of the discrete lattice action ~\cite{Agishtein:1992xx} and can be seen as follows, 
\bea\label{eq:shiftsym}
	S_{\mathrm{lat}} &=& \sum_{\langle \xy \rangle} \bigg ((\Phi_{x} - \Phi_{y})^{2} + \sum_{x} m_{0}^{2} \Phi_{x}^{2} \bigg)  \\
 &=&   \sum_{\langle \xy \rangle} \bigg ((\text{D} + 1 + m_{0}^{2})\delta_{xy} - C_{xy} \bigg)\Phi_{\x}\Phi_{\y}          
 \eea
 where $C_{\xy}$ is the simplex neighbor (or connectivity) matrix, $m_{0}$ is the bare 
 mass and $D$ is the space-time dimension. 
 For zero bare mass, there is a shift symmetry, $\Phi \to \Phi + c$,
which implies that the renormalized mass should vanish in the limit that the bare mass is sent to zero. 

\section{Computation and results}

We restrict our analysis to the ensembles close to the critical line of the phase transition, where it was shown that the 
geometry has semi-classical properties, and on each degenerate dynamical triangulation, 
we calculate the propagator using, 
\begin{equation}
G_{\xy} = (-\B + m_{0}^{2})^{-1}_{\xy} 
\end{equation} 
The definition of the discrete Laplacian is given by, 

\begin{figure}[htbp]
  \centering
  \includegraphics[height=4.76cm]{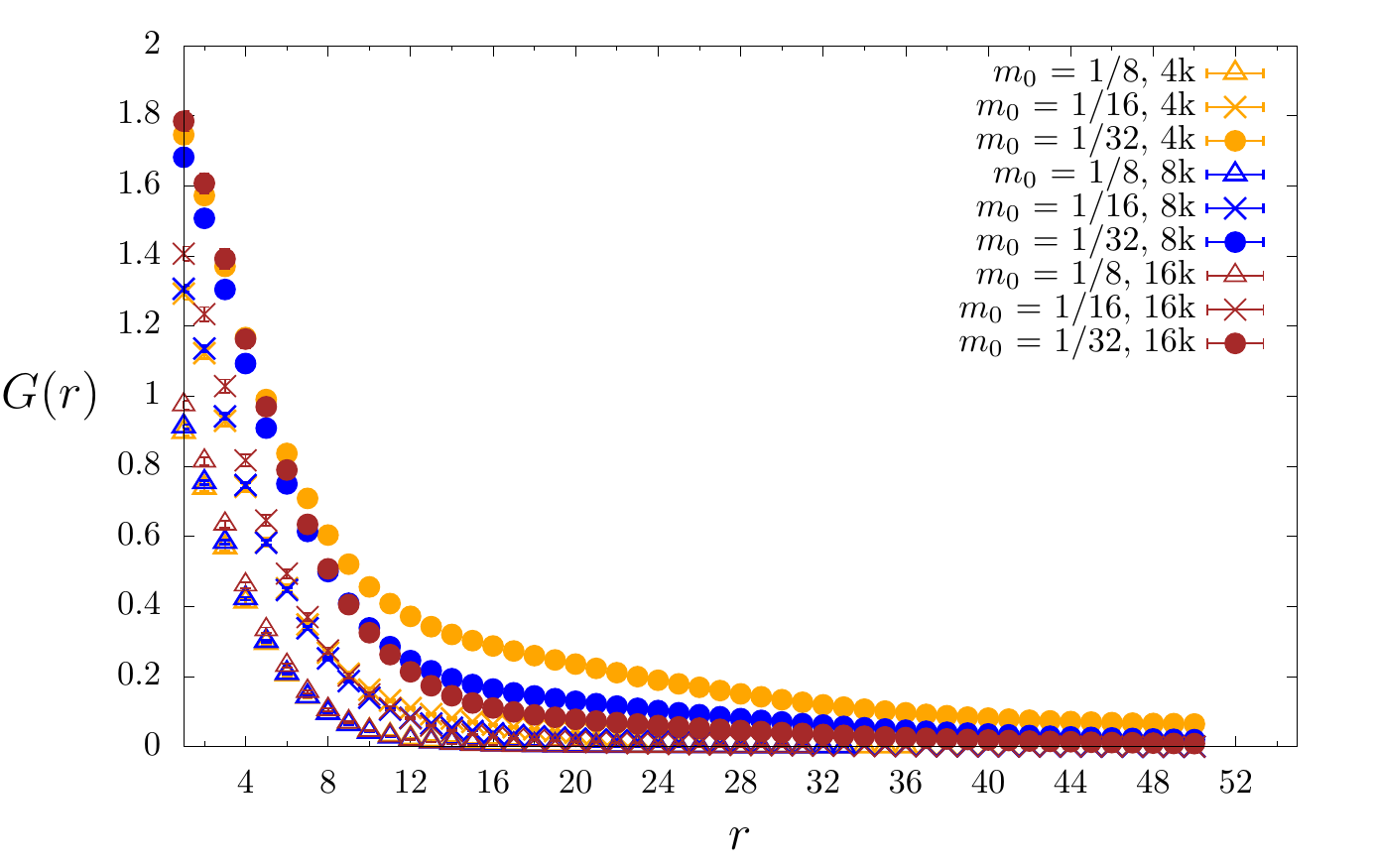}\hfill \includegraphics[height=4.76cm]{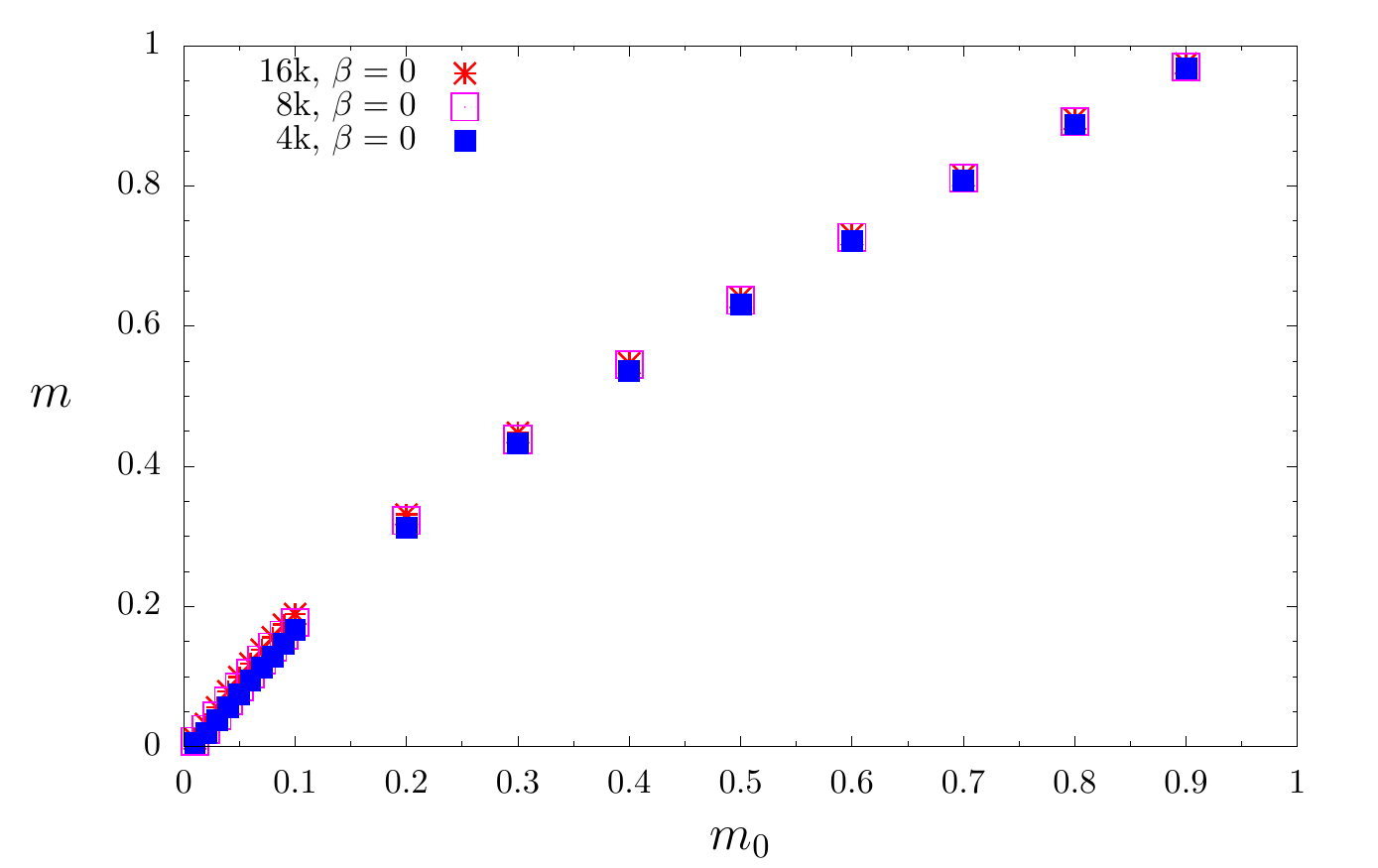} \\[6 pt]
  \includegraphics[height=4.76cm]{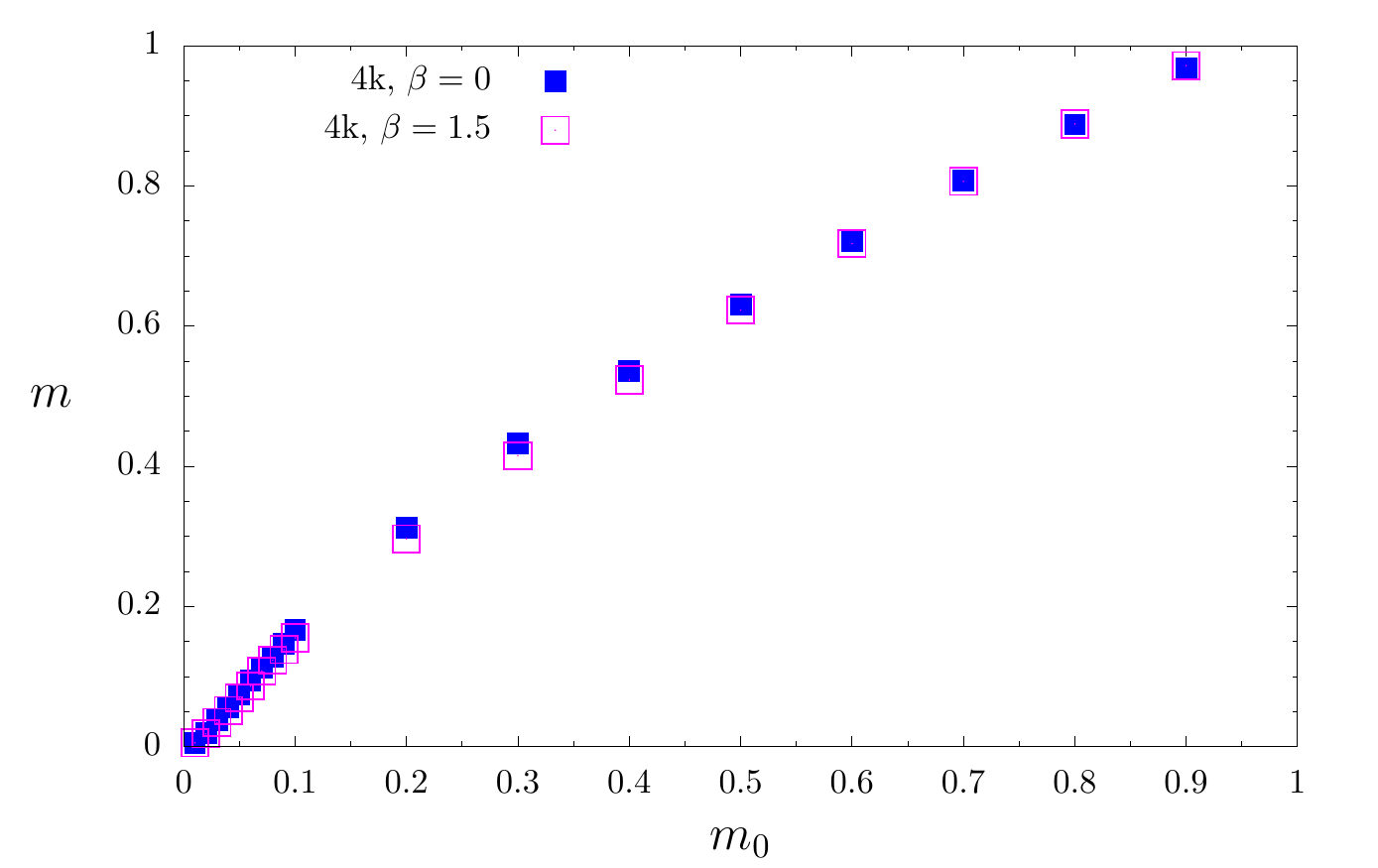}
  \caption{\label{fig:beta0prop}\textbf{Left:} $G(r)$ for different volumes at $\beta=0$. \textbf{Right:} The renormalized mass versus the bare mass for different lattice volumes ($i.e. ~ N_{4}$) with $\beta=0$. \textbf{Bottom:} The $\beta$ 
  dependence on a fixed lattice volume. The range used for fitting the data $r \in (24,28)$ for $16$k, $r \in (19,23)$ for 8k and $r \in (16,20)$ for 4k respectively.}
\end{figure}
\begin{equation}
-\B_{\xy} =   \hspace{6mm} \begin{cases}
   D+1 \hspace{5mm} \text{if x = y}  \quad \phantom{0}  \\
    -1 \hspace{10mm}    \text{if x \& y are nearest neighbors}   \\
     0 \hspace{13mm}      \text{otherwise} 
      \end{cases}
    \end{equation} 
Here, $D+1$ is the coordination number of a $D$-simplex, which is 5 in our case. Unlike 
the combinatorial case studied in \cite{deBakker:1996qf, deBakker:1995ed}, here we study degenerate triangulations.
In such a setting, a given four-simplex can have non-unique neighbors. This enables us to 
construct the laplacian for configurations such that the sum of any particular row is just $ m_{0}^{2}$. In the limit 
of vanishing bare mass ($ m_{0} \to 0$), we have an exact zero mode of the operator corresponding to the
zero eigenvalue of the Laplacian. It should be noted that there exists an alternate 
geometric description of the laplacian in terms of boundary ($\partial$) \footnote{The action of $\partial$ on
a given $\textit{p-simplex}$ yields an directed sum of (p - 1)-simplices on its boundary, whereas, $\overline{\partial}$ yields
directed sum of (p + 1)-simplices} and
co-boundary operators ($\overline{\partial}$). We can then write the laplacian as, 
$ \B = -\partial \overline{\partial} - \overline{\partial} \partial  = (\partial -  \overline{\partial})^{2}$, where
we have used $\partial^{2} = \overline{\partial}^{2} = 0 $. See ~\cite{Catterall:2018lkj} for details and 
its role in gravitational systems coupled to fermions. 

Similar to the continuum definition of the massive lattice scalar propagator, we can define the same quantity 
on the lattice as, 
    
  \bea\label{eq:latticeprop}
  G_{\xy}(r)&=&  \bigg[\frac{1}{(-\B + m_{0}^{2})} \bigg]_{\xy}  \\  
  &\propto & e^{-mr}       
 \eea
  where, $m$ is the renormalized mass. 
 Unlike the previous study \cite{deBakker:1996qf}, we have not used smeared sources for the inversion of the matrix. 
 However, it might be crucial for the future calculation of binding energy. 
 In calculating the propagator over all configurations, we have considered $\emph{five}$
 sources ($N_{\text{src}} = 5$) per configuration and checked that our results are unaffected 
 by this choice. 

\begin{table}
\begin{center}
\begin{tabular}{ccccccc}
\hline \hline
$a_{\rm rel}$ & $\beta$ & $\kappa_2$ & $N_4$ &  $\text{Physical volume}$ \\
\hline
1.47(10) & 1.5 & 0.5886 & \ \ 4000  & 0.054  \\
1 & 0.0 & 1.669 &         \ \ 4000 & 0.25  \\
1 & 0.0 & 1.7024 &     \ \  8000 & 0.125  \\
1 & 0.0 & 1.7325 &     \ \ 16000 & 0.0625  \\
\hline
\hline

\end{tabular}
\caption{\label{table1}The ensembles used in this work. 
There are several other ensembles that we have generated but are not included in this proceedings. 
See ~\cite{Laiho:2016nlp} for details. }
\end{center}
\end{table}

\subsection{Binding energy and Mass anomalous dimension of scalar fields}
The calculation of the one-particle and two-particle propagators on generated ensembles can be fitted to 
the form as given below

\begin{eqnarray}
G(r) &=& A r^{\alpha} e^{-mr}  \\
G^{(2)}(r) &=& B r^{\beta} e^{-Mr}
\end{eqnarray}
where $M$ is the energy of the two-particle state. The coefficient of the fits can then be used to calculate the 
binding energy associated as

\begin{eqnarray}
E_{b}(r) &=& \frac{1}{r} ~\text{ln} \Bigg(\frac{G(r)^{2}}{G^{(2)}(r)}\Bigg) 
\end{eqnarray}
such that at long distances, \emph{i.e.} $ r \to \infty$, the negative value of the binding energy, $E_{b}(r \to \infty) = M - 2m$ would represent an 
attractive force.   

We now discuss the calculation of the mass anomalous dimension from a determination of the mass renormalization factor at the 
same physical volume but different lattice spacings. The bare and renormalized lagrangians
are  

\begin{align} & L_{b}=\dfrac {1} {2}\partial _{\mu}\varphi_{0} \partial ^{\mu }\varphi_{0}  -\dfrac {1}{2} m_{0}^{2}\varphi_{0} ^{2}\end{align}
\begin{align} & L_{r}=\dfrac {1} {2} Z_{\varphi} \partial _{\mu}\varphi \partial ^{\mu }\varphi  -\dfrac {1}{2} Z_{m} m^{2}\varphi ^{2}.\end{align}
Comparing them we get, 
\beq
\varphi _{0}\left( x\right) =Z_{\varphi}^{1/2} \varphi \left( x\right)
\eeq and,  
\beq
\label{eq:Zm1}
m = \sqrt{\frac{Z_{\varphi}}{Z_{m}}} ~ m_{0} 
\eeq
Taking the log of (~\ref{eq:Zm1}) and then taking the derivative with respect to $\ln \mu$ and identifying $\mu = 1/a$, we get: 

\beq
\frac{d\ln (m)}{d \ln a} = \dfrac {1} {2} \Big( \frac{d \ln F}{d \ln a} \Big)
\eeq
where we have used the shorthand $ F = Z_{\varphi} / Z_{m} $. Then, the mass anomalous dimension
is given by, 

\beq
\gamma_{m} = - \dfrac {1} {2} \Big( \frac{d \ln F}{d \ln a} \Big)
\eeq
In the future, we will explore additional volumes and lattice spacings and report on the results for gravitational binding energy and 
anomalous dimensions.

\vspace{8 pt}
\noindent {\textbf{Acknowledgments:}}~ 
This research was supported by the US Department of Energy (DOE), Office of Science, Office of High Energy Physics, 
under Award Number DE-SC0009998.

\bibliographystyle{utphys}
\bibliography{lattice18.bib}
\end{document}